\documentstyle[12pt]{article}

\newcommand{\EQ}{\begin{equation}}
\newcommand{\EN}{\end{equation}}
\newcommand{\bea}{\begin{eqnarray}}
\newcommand{\ena}{\end{eqnarray}}
\newcommand{\bdis}{\begin{displaymath}}
\newcommand{\edis}{\end{displaymath}}
\newcommand{\vs}[1]{\vspace{#1 mm}}

\renewcommand{\a}{\alpha}

\renewcommand{\d}{\delta}
\renewcommand{\v}{\Delta}

\renewcommand{\t}{\tau}

\newcommand{\pa}{\partial}

\newcommand{\nn}{\nonumber \\}

\begin{document}

\topmargin 0pt
\oddsidemargin 5mm

\newcommand{\NP}[1]{Nucl.\ Phys.\ {\bf #1}}
\newcommand{\PL}[1]{Phys.\ Lett.\ {\bf #1}}
\newcommand{\CMP}[1]{Comm.\ Math.\ Phys.\ {\bf #1}}
\newcommand{\PR}[1]{Phys.\ Rev.\ {\bf #1}}
\newcommand{\PRL}[1]{Phys.\ Rev.\ Lett.\ {\bf #1}}
\newcommand{\PREP}[1]{Phys.\ Rep.\ {\bf #1}}
\newcommand{\PTP}[1]{Prog.\ Theor.\ Phys.\ {\bf #1}}
\newcommand{\PTPS}[1]{Prog.\ Theor.\ Phys.\ Suppl.\ {\bf #1}}
\newcommand{\NC}[1]{Nuovo.\ Cim.\ {\bf #1}}
\newcommand{\JPSJ}[1]{J.\ Phys.\ Soc.\ Japan\ {\bf #1}}
\newcommand{\MPL}[1]{Mod.\ Phys.\ Lett.\ {\bf #1}}
\newcommand{\IJMP}[1]{Int.\ Jour.\ Mod.\ Phys.\ {\bf #1}}
\newcommand{\AP}[1]{Ann.\ Phys.\ {\bf #1}}
\newcommand{\RMP}[1]{Rev.\ Mod.\ Phys.\ {\bf #1}}
\newcommand{\PMI}[1]{Publ.\ Math.\ IHES\ {\bf #1}}
\newcommand{\JETP}[1]{Sov.\ Phys.\ J.E.T.P.\ {\bf #1}}
\newcommand{\TOP}[1]{Topology\ {\bf #1}}
\newcommand{\AM}[1]{Ann.\ Math.\ {\bf #1}}
\newcommand{\LMP}[1]{Lett.\ Math.\ Phys.\ {\bf #1}}
\newcommand{\CRASP}[1]{C.R.\ Acad.\ Sci.\ Paris\ {\bf #1}}
\newcommand{\JDG}[1]{J.\ Diff.\ Geom.\ {\bf #1}}
\newcommand{\JSP}[1]{J.\ Stat.\ Phys.\ {\bf #1}}

\begin{titlepage}
\setcounter{page}{0}
\begin{flushright}
NBI-HE-94-54\\
December 1994\\
hep-th/9411232\\
\end{flushright}

\vs{8}
\begin{center}
{\Large On Field Theories of Loops}

\vs{15}
{\large Naohito Nakazawa\footnote{e-mail address:
nakazawa@nbivax.nbi.dk, nakazawa@ps1.yukawa.kyoto-u.ac.jp}}\\
{\em The Niels Bohr Institute, Blegdamsvej 17, DK-2100 Copenhagen \O,
Denmark \\
and \\
Department of Physics, Shimane University,
Matsue 690, Japan}
\end{center}

\vs{8}
\centerline{{\bf{Abstract}}}

We apply stochastic quantization method to
real symmetric matrix models for the second quantization of
non-orientable loops in both discretized and continuum levels.
The stochastic process defined by the
Langevin equation in loop space describes the time evolution of the
non-orientable loops defined on non-orientable 2D surfaces.
The corresponding Fokker-Planck
hamiltonian deduces a non-orientable string field theory
at the continuum limit.

\end{titlepage}
\newpage
\renewcommand{\thefootnote}{\arabic{footnote}}
\setcounter{footnote}{0}

String field theory~\cite{SFT} is believed to be the most promising approach
to investigate non-perturbative effect in string theories.
Recently, non-critical string field theories have been proposed
for $c = 0$~\cite{IK}\cite{JR}\cite{Wata} and for $0< c < 1$~\cite{IIKMNS}.
Among these, some~\cite{IK}\cite{Wata}\cite{IIKMNS}
are based on the transfer-matrix formalism~\cite{KKMW} in dynamical
triangulation of random surfaces~\cite{DT}.
While the other~\cite{JR} is derived by using
stochastic quantization method~\cite{PW}.

In the approach by stochastic quantization,
introducing the loop variable
${\rm tr} M^n$~\cite{JS} for hermitian matrix models and interpreting the
fictitious time as a time coordinate,
Jevicki and Rodrigues~\cite{JR} showed that the
Fokker-Planck hamiltonian ( or loop space hamiltonian ),
in which the loop variable and the conjugate
momentum are identified with the creation operator and the annihilation
operator respectively, realizes the string field theories
which are equivalent to the field theory derived
by Ishibashi-Kawai~\cite{IK}.
Inspired by the work in Ref.~\cite{JR}, we apply
stochastic quantization method to real symmetric matrix models~\cite{RS}
and show that it leads to a field theory of non-orientable (non-critical)
strings.
The stochastic process defined by the
Langevin equation in loop space describes the time evolution of the
non-orientable loops on non-orientable 2D surfaces.
The corresponding Fokker-Planck
hamiltonian is a loop space hamiltonian of
non-orientable string field theories. At the equilibrium limit, it deduces
the Virasoro constraint equation for the probability distribution
functional.
The continuum limit of the field theory of discretized non-orientable loops
is taken for the simplest one-matrix case ( $c = 0$ )
and deduces the continuum field theory of
non-orientable strings.


Let us start with the Langevin equation for one matrix model,
\bea
{\v}M_{ij}(\t)
&=&  - {\pa \over\pa M}S(M)_{ij}(\t) \v\t + \v\xi_{ij}(\t)        \ , \nn
S(M)
&=& - \sum_{\a = 0} {g_\a \over \a + 2} N^{- \a /2}{\rm tr}M^{\a + 2}  \ , \nn
\ena
$M_{ij}$ denotes a real symmetric matrix.
The fictitious time $\t$ is
discretized with the unit time step $\v\t$. We consider the
discretized version of time evolution
$
M_{ij} ( \t+\v\t ) \equiv M_{ij} ( \t ) + \v M_{ij}( \t ) \ ,
$
with the Langevin equation for convenience of
stochastic calculus and for understanding the corresponding
stochastic process precisely.
The discretized fictitious time development with $\v\t$ precisely
corresponds to the
one step deformation in dynamical triangulation in random surfaces.
In the following argument,
the specific form of the action of the matrix model is not relevant.
The correlation of the white noise $\v\xi_{ij}$ is
defined by
\EQ
<\v\xi_{ij}(\t) \v\xi_{kl}(\t)>_\xi
= \v\t \big( \d_{il} \d_{jk} + \d_{ik} \d_{jl} \big)   \ .
\EN
It is uniquely determined\footnote{
For an hermitian matrix $M_{ij}$ in (1),
the nose correlation is
$
<\v\xi_{ij}(\t) \v\xi_{kl}(\t)>_\xi
= 2 \v\t \d_{il} \d_{jk}                        \ .
$
%
} from the requirement that
(1) is transformed covariantly preserving the white noise
correlation (2) invariant under the transformation
$
M \rightarrow U M U^{-1}   \ ,
$
where $U$ denotes orthogonal matrices for the
real symmetric matrix models.

The basic field variables are loop variables
$
\phi_n = {\rm tr}(M^n) N^{-1 - {n\over 2}}   \ .
$
Following to Ito's stochastic calculus~\cite{I}, we evaluate
\bea
\v\phi_n
&\equiv & \phi_n (\t + \v\t) - \phi_n (\t)                  \ , \nn
&=& n {\rm tr}(\v M M^{n-1}) N^{-1 - {n\over 2}}
+ {1\over 2} n \sum_{k=0}^{n-2} {\rm tr}(\v M M^{k} \v M M^{n-k-2})
N^{-1 - {n\over 2}}
+ O(\v\t^{3/2})                          \  .   \nn
\ena
The terms in R.H.S. should be of the order $\v\t$, thus we obtain
\bea
\v\phi_n
&=& \v\t { n\over 2} \big\{ \sum_{k=0}^{n-2}
\phi_k \phi_{n-k-2}
+ (n-1) {1\over N} \phi_{n-2}                 \big\}
+ \v\t\ n \sum_{\a=0} g_\a \phi_{n+\a}   +   \v \zeta_{n-1}    \ ,   \nn
\v\zeta_{n-1}
&\equiv& n {\rm tr}(\v\xi M^{n-1}) N^{-1 - {n\over 2}}     \ .  \nn
\ena
The correlation of the new noise variables appeared in (4) is given by
\EQ
<\v\zeta_{m-1} (\t) \v\zeta_{n-1} (\t)>_\xi
= \v\t {2\over N^2} n m < \phi_{m+n-2} (\t) >_\xi   ,
\EN
The new noise is not a simple white noise but includes the
value of the loop variable itself.
In a practical sense, it might be
tedious to generate the noise variable.
We notice that
$\phi_{m+n-2}(\t)$ in R.H.S. of eq.(5) does not include the white noise
at $\t$ but the series of noises up to
the one step (fictitious time unit $\v\t$) before.
This means that the expectation value in R.H.S. should be defined with
respect to the white noise correlation up to the fictitious time $\t - \v\t$.

We also notice
$
<\v\zeta_n (\t)>_\xi = 0
$
by means of Ito's stochastic calculus.
In the context of SQM approach,
the property of the noise yields
the Schwinger-Dyson equation by assuming the existence of the
equilibrium limit at the infinite fictitious time, or equivalently,
$
\lim_{\t \rightarrow \infty} < \v \phi_n (\t) >_\xi = 0    \  .
$
We have,
\EQ
< {n\over 2} \sum_{k=0}^{n-2}
\phi_k \phi_{n-k-2}
+ {1\over 2}(n-1) {1\over N} \phi_{n-2}
+ n \sum_{\a=0} g_\a \phi_{n+\a}     >_\xi = 0                \ .
\EN

The order of the noise correlation (5), ${1/N^2}$,
realizes the factorization condition in the large N limit~\cite{GH}.
Therefore we obtain the S-D equation at large N limit for discretized
non-orientable strings.
\EQ
{1\over 2}\sum_{k=0}^{n-2} < \phi_k >_\xi< \phi_{n-k-2} >_\xi
+ \sum_{\a = 0} g_\a < \phi_{n + \a}  >_\xi = 0                \ .
\EN
This shows that the S-D equation for non-orientable strings takes the
same form as that for orientable strings at large N limit.
The correspondence at the large N limit is exact if we define the
corresponding hermitian matrix model by replacing all the couplings
, $g_\a \rightarrow 2g_\a$ in (1).
As a consequence,
the disc amplitude in non-orientable strings is exactly the same as
that in orientable
strings.

The geometrical meaning of the stochastic process described by
the Langevin equation (4) is the following.
The one step fictitious time evolution
of a discretized loop,
%
$
\phi_n (\t) \rightarrow \phi_n(\t) + \v \phi_n (\t)    \ ,
$
generates the splitting of the loop into two smaller pieces,
$\phi_k$ and
$\phi_{n-k-2}$. The process is described by the first term in R.H.S.
of (4). In a field theoretical sense, it is interpreted as
the annihilation of the loop $\phi_n$ and the
simultaneous pair creation of loops, $\phi_k$ and $\phi_{n-k-2}$.
The first term in R.H.S. of (4) preserves the
orientation of these loops, while the second term, which
is the characteristic term of the
order of ${1\over N}$
for non-orientable interaction,
does not preserve the orientation.
Since the new noise variables in (5),
$\v\zeta_{n-1}$'s, are translated to \lq\lq annihilation" operators in
the corresponding Fokker-Planck hamiltonian,
the factor 2 in the correlation (5) for the
new noise variables comes from the sum of the orientation preserving and
non-preserving merging interactions. Namely, (5) describes the
simultaneous annihilation of two loops
$\phi_{m}$ and $\phi_{n}$ and the creation of a loop $\phi_{m+n-2}$.
The geometrical picture allows us to identify the power \lq\lq $n$" of matrices
in $\phi_n$ to the length of the discretized non-orientable loop $\phi_n$.
We notice that, in each time step, the interaction process decreases the
discretized loop length by the unit \lq\lq 2".
The process which comes from the
original action of matrix models extends the length
of discretized loops.
These features are
equivalent to the transfer-matrix formalism~\cite{KKMW}
in dynamical triangulation of 2D
random surfaces in which the one step deformation of a specified loop on
a triangulated surface defines a discretized (proper) time evolution
of the loop.

The definition of the F-P hamiltonian operator
gives the precise definition
of a field theory of second quantized non-orientable strings.
In terms of the expectation value of an observable $O(\phi)$, a
function of $\phi_n$'s, the F-P hamiltonian operator ${\hat H}_{FP}$ is
defined by,
\EQ
<\phi (0)| {\rm e}^{- \t {\hat H}_{FP} } O({\hat \phi})|0>
\equiv <O( \phi_\xi(\t) )>_\xi                  \  .
\EN
In R.H.S., $\phi_\xi(\t)$ denotes the solution of the Langevin equation (4)
with the initial configuration $\phi (0) \neq 0$.
The time evolution of R.H.S. is given by,
\bea
<\v O(\phi(\t))>_\xi
&=& <\sum_m \pa_m O (\phi(\t)) \v \phi_m
+ {1\over 2}\sum_{m,n} \pa_m \pa_n O(\phi(\t)) \v \phi_m \v \phi_n >_\xi
+ O(\v \t^{3/2})         \ ,      \nn
&\equiv& - \v \t < H_{FP}(\t) O(\phi(\t)) >_\xi      \ , \nn
\ena
where $\pa_n \equiv {\pa \over \pa \phi_n}$. By substituting the Langevin
equation (4) and the noise correlation (5) into (9), we obtain
\bea
H_{FP}(\t)
&=& - \sum_{n>0}X_{n} n\pi_n     \ , \nn
X_n
&\equiv&
{1\over N^2} \sum_m m\phi_{m+n-2}\pi_m
+ {1\over 2}\sum_{k=0}^{n-2}\phi_k\phi_{n-k-2}
+ {1\over 2}(n-1) {1\over N} \phi_{n-2}
+ \sum_{\a =0}g_\a \phi_{\a + 2}                     \nn  ,
\ena
where
$
\pi_n \equiv {\pa \over\pa \phi_n}     \ .
$
To define the operator formalism corresponding to eq.(8),
we introduce
${\hat \phi}_m$ and ${\hat \pi}_m$ as the creation and the annihilation
operators for the loop with the length $n$, respectively. Then we assume
the commutation relation
$
[ {\hat \pi}_m  ,  {\hat \phi}_n ]
= \d_{mn}                 \ ,
$
and the existence of the vacuum, $|0>$, with
$
{\hat \pi}_m |0> =  <0|{\hat \phi}_m
= 0 \quad {\rm for} \  m > 0      \  .
$
In the representation,
$
<Q| \equiv <0|{\rm e}^{\sum_m Q_m {\hat \pi}_m}  \
$
and
$
|Q> \equiv \Pi_m \d ({\hat \phi}_m - Q_m ) |0>  \ ,
$
the F-P hamiltonian operator
${\hat H}_{FP}$ in (8) is given by replacing
$\phi_m \rightarrow {\hat \phi}_m$, and
$\pi_m \rightarrow {\hat \pi}_m$ in $H_{FP}$ in (10) with the same operator
ordering.

%
%

{}From the equality (8), the probability distribution function $P(\phi, \t)$
, which is defined by
$
<O( \phi(\t) )>_\xi
\equiv \int \!\Pi_{n} d\phi_n O(\phi) P(\phi, \t)    \ ,
$
is given by,
\EQ
P(\phi,\t)
= <\phi (0)| {\rm e}^{- \t {\hat H}_{FP} } | \phi >          \  .
\EN
The initial distribution,
$
P(\phi, 0) = \Pi_m \d (\phi_m - \phi_m (0) )     \ ,
$
represents the initial value of the solution of the Langevin equation
(4). Eq.(11) follows the Fokker-Planck equation for the probability
distribution,
\EQ
\v P(\phi,\t)
= + \v\t {\tilde H}_{FP}P(\phi,\t)      \ ,
\EN
where ${\tilde H}_{FP}$ is the adjoint of $H_{FP}$ in (10),
\bea
{\tilde H}_{FP}(\t)
&=& - n\pi_n \sum_{n>0}{\tilde X}_{n}      \ , \nn
{\tilde X}_n
\equiv
&-& {1\over N^2} \sum_m m\pi_m \phi_{m+n-2}
+ {1\over 2}\sum_{k=0}^{n-2}\phi_k\phi_{n-k-2}
+ {1\over 2}(n-1) {1\over N} \phi_{n-2}                 \nn
&+& \sum_{\a =0}g_\a \phi_{\a + 2}
                  \  .        \nn
\ena
%

In the context of stochastic quantization, the F-P hamiltonian (10)
in loop space was found for
hermitian matrix models. The remarkable observation
was that it includes the Virasoro constraint~\cite{JR}.
Since the fictitious time evolution is generated by the noise essentially,
the emergence of Virasoro constraint
is traced to the noise correlations in eq.(5) which
are equivalent to the insertion of matrices
into the loop variable,
$
M \rightarrow M + \v\t M^{m-1}           \ ,
$
in $\phi_n$. It generates the transformation
$
[ - \v\t L_{m-2} , \phi_n ] = n \v\t \phi_{m+n-2}   \ ,
$
which corresponds to the noise correlation (5).
In fact,
for real symmetric matrix models ( non-orientable strings ),
$L_n \equiv - {N^2} X_{n+2}$
satisfies the Virasoro algebra without central extension,
\EQ
[ L_m, L_n ] = (m-n)L_{m+n}      .
\EN
We notice that, although ${\hat H}_{FP}$ is not hermitian,
one can define an hermitian F-P hamiltonian from ${\hat H}_{FP}$
as a direct consequence of the fact that Ito's stochastic calculus
automatically picks up the Jacobian factor~\cite{Na}
( or more precisely, invariant
measure ) which is induced in the space of
loop variables by the change of dynamical
variables from a matrix to loop variables~\cite{JS}.

It is also worthwhile to note that
the F-P equation (12) realizes the Virasoro constraint for the
probability distribution. Namely,
${\tilde L}_n \equiv {N^2} {\tilde X}_{n+2}$
also satisfies the Virasoro algebra without central extension (14).
Therefore,
the F-P equation deduces a constraint equation for the
distribution function even at the
discretized level, justifying the generation of the partition function
which satisfies the Virasoro constraint at the infinite fictitious time.
\EQ
{\tilde L}_n \lim_{\t \rightarrow \infty}P(\phi, \t)  = 0
\ , \quad {\rm for } \
n = -1, 0, 1, ...                  \ . \nn
\EN
For hermitian matrix models, the Virasoro constraint
for the partition function (15)
was found as the S-D equation~\cite{AJM}.
In the continuum limit, it deduces
the continuum version of the Virasoro
constraints~\cite{FKN}.
The expressions (8) and (11) also give a constraint on possible initial
condition dependence of the expectation value and the partition
function at the infinite fictitious time limit, such as,
$
\lim_{\t \rightarrow \infty}H_{FP}[\phi(0), {\pa\over\pa \phi(0)}]
P[\phi, \t] = 0        \ .
$
This implies that these quantities may have the initial
value dependence up to the solution of the Virasoro constraint.

Now we take the continuum limit. First we introduce a length scale
\lq\lq $a$ " and define the physical length of the loop created by
$\phi_n$ with
$
l = n a
$. Then we may redefine field variables and the fictitious time
at the continuum limit as follows.
\bea
G_{st}
&\equiv& N^{-2} a^{- D}       \  ,\nn
d\t
&\equiv& a^{- 2 + D/2} \v\t       \  ,\nn
\Phi (l)
&\equiv& a^{- D/2 } \phi_n      \ , \nn
\Pi (l)
&\equiv& a^{- 1 + D/2 } \pi_n      \ ,\nn
\ena
where we would like to keep the string coupling constant, $G_{st}$, finite at
the double scaling limit~\cite{DS}.
For the existence of the smooth
limit from the discretized fictitious time evolution to the \lq\lq
continuum" one, we require the condition, ${D\over 2} - 2 > 0$.
The scaling dimensions of all the quantities in (16)
have been
determined except the scaling dimension of the string coupling
constant, $D$,
by assuming~\cite{IK}\cite{JR},
\bea
\v\t H_{FP}
&=& d\t {\cal H}_{FP}          \  ,  \nn
\big[ \Pi(l) ,  \Phi(l)  \big]
&=& \d (l - l')                \  . \nn
\ena
Then we obtain the continuum F-P hamiltonian, ${\cal H}_{FP}$, from
$H_{FP}$ at the continuum limit,
\bea
{\cal H}^{non-or.}_{FP}
&=& - {1\over 2}\int_0^{\infty}\!dl
\big\{ 2 G_{st}\int_0^{\infty}\!dl' \Phi(l+l')l'\Pi(l')l\Pi(l) +
\int_0^{l}\!dl' \Phi(l-l')\Phi(l') l\Pi(l)                   \nn
&{}& \quad + \sqrt{G_{st}} l\Phi(l) l\Pi(l)
 + \rho(l)\Pi(l)       \big\}             \ , \nn
\ena
for non-orientable strings.
By the redefinition (16), the F-P hamiltonian (18) is uniquely fixed
at the continuum limit except the cosmological term.
To specify the explicit
form of the cosmological term $\rho(l)$ in (18),
we have to carefully evaluate the contribution which
comes from the 3-point splitting interaction term and the matrix model
potential,
$
a^{1-D}\sum_{\a=0}g_{\a}\phi_{n + \a}       \  ,
$
at the double scaling limit of the real symmetric matrix model.
Here we remember that
the S-D equation for non-orientable string at large N limit takes exactly
the same form as the
orientable one under the suitable choice of the matrix model
coupling constants.
Since the continuum limit is taken
by using the universal part of the disc amplitude~\cite{IK}, we naively expect
$\rho (l)$ takes the same form as that for orientable strings.

To show explicitly this is indeed the case,
we take the continuum limit in the real symmetric matrix model
with the same procedure in Ref.~\cite{JR}.
We consider the simplest matrix potential given by,
$g_0 = -1/2,\ g_1 = g/2, g_2 = g_3 =... = 0$ in (1),
which corresponds to $c=0$.
Let us introduce the string field variable
\bea
\phi (z)
&\equiv& \sum_n z^{-1-n}\phi_n
= {1\over N}{\rm tr}{1\over z - M N^{-1/2}}           \ , \nn
\v\zeta (z)
&\equiv& \sum_n z^{-1-n} \v\zeta_n                       \ . \nn
\ena
To take the continuum limit, we redefine the field variable,
\bea
\phi (z) &\equiv& {1\over 2}(z - g z^2 )
+ c_0 z_c^{-1} a^{3/2}\Phi(u)    \ , \nn
\v\zeta (z) &\equiv& c_0 z_c^{-1} a^{3/2} d{{\tilde \zeta}} (u)
\ , \nn
\ena
where we have introduced the \lq\lq renormalized" parameters,
$
z \equiv z_c {\rm e }^{a u}       \  ,
$
and
$
g \equiv g_c {\rm e}^{ - c_1 a^2  t }    \  ,
$
and the \lq\lq continuum" fictitious time
$
d\t \equiv c_0 z_c^{-2} a^{1/2} \v\t                \ ,
$
where $z_c = 3^{1\over 4}(3^{1\over 2} + 1)$ and
$g_c = {1\over 2 \cdot 3^{3\over 4}}$ are the critical values
and $t$ denotes the cosmological
constant. The constants $c_0$ and $c_1$
are chosen for convenience.
The scaling dimension of all the quantities have been
determined so that the string coupling,
$
1/G_{st} \equiv c_0^2 N^2 a^5
$
, is fixed at the double scaling limit~\cite{DS}.
Then we have the following Langevin equation in continuum limit.
\bea
d\Phi(u)
&=& d{ {\tilde \zeta}} (u)
- {1\over 2}d\t \pa_u \big\{ \Phi(u)^2
- {1\over N a^{5/2}} \pa_u \big( \Phi(u) \big)
- a^{-3}\big( {1\over 4} ( z - g z^2)^2 + g z
    \big) \big\}               \ , \nn
&=& d{ {\tilde \zeta}} (u)
- {1\over 2}d\t \big\{ \pa_u \Phi(u)^2
- {\tilde \rho}(u)
- {1\over N a^{5/2}} \pa_u^2 \big( \Phi(u) \big)    \big\}  + O(a)   \ , \nn
<d{ {\tilde \zeta}} (u) d{ {\tilde \zeta}} (u')>
&=& {2 d\t\over N^2 a^5}
\pa_u \pa_{u'}\big\{ {1\over u'- u}\big( \Phi(u) - \Phi(u') \big) \big\}
+ O(a) \ , \nn
{\tilde \rho}(u)
&=& 3 u^2 - {3 t\over 4}                    \ .\nn
\ena
We have only picked up the terms in the noise correlation which survive at the
continuum limit. We notice, since
$
\pa_u^2 ( z - g z^2 ) = O (a^2)
$, the field redefinition in (20) is irrelevant in the term of the order
$1/( N a^{5/2} )$. ${\tilde \rho}(u)$ in (21) is precisely
the cosmological term
appeared in the orientable string.

By the Laplace transformation,
\EQ
\Phi(u) = \int_{0}^{\infty}\!dl{\rm e}^{-u l}\Phi(l)    \ ,
\EN
we obtain the Langevin
equation which is equivalent to the continuum F-P hamiltonian (18),
\bea
d\Phi(l)
&=& d\zeta (l)
+ {1\over 2}d\t \big\{ l \int_{0}^{l}\!dl' \Phi(l') \Phi (l-l')
+ \rho (l) + \sqrt{G_{st}} l^2 \Phi(l)
\big\}                             \ , \nn
<d\zeta (l) d\zeta (l')>
&=& 2 d\t G_{st}
l l' < \Phi(l + l') >              \ , \nn
\rho(l)
&=& 3 \d''(l) - {3 t\over 4}\d (l)         \ , \nn
\ena
for non-orientable string. It is consistent with the
naive continuum limit of its discretized version (4) except
the term $\rho (l)$.
By the same procedure, the Langevin equation for orientable string is given
by
\bea
d\Phi(l)
&=& d\zeta (l)
+ d\t \big\{ l \int_{0}^{l}\!dl' \Phi(l') \Phi (l-l')
+ \rho (l)   \big\}                      \ , \nn
<d\zeta (l) d\zeta (l')>
&=& 2 d\t G_{st}
l l' < \Phi(l + l') > \ , \nn
\ena
in the hermitian matrix model with replacing the coupling constants,
$g_\a \rightarrow 2 g_\a$, in (1).
As we have shown explicitly, the cosmological term
$\rho (l)$ takes the same form both for orientable and non-orientable strings.
The field theory of non-orientable strings
is also consistent with ref.~\cite{Wata} in transfer matrix formalism.
The double scaling limit of the real symmetric matrix model
has been studied in a
quartic potential~\cite{RS}, while our result shows
that it happens in the cubic
potential as well. We notice that the continuum
F-P hamiltonian includes the continuum Virasoro generator ${\cal L}(l)$,
\bea
{\cal L}(l)
&=& -
\big\{ \int_0^{\infty}\!dl' \Phi(l+l')l'\Pi(l') +
{1\over 2 G_{st}}\int_0^{l}\!dl' \Phi(l-l')\Phi(l')                    \nn
&{}& \quad + {1\over 2 \sqrt{G_{st}}} l\Phi(l)
 + {1\over 2 G_{st}}{\rho(l)\over l}       \big\}             \ . \nn
\ena
These generators satisfy the continuum Virasoro algebra,
$
[{\cal L}(l), {\cal L}(l')]
= (l-l'){\cal L} (l+l')      \ .
$

Let us briefly comment on the multi-matrix model cases.
We may start from a set of Langevin equations.
\bea
\v M(p)_{ij}(\t)
&=&  - {\pa \over\pa M(p)}S(M)_{ij}(\t) \v\t +
\v\xi_{ij}(\t)                         \  ,     \nn
<\v\xi(p)_{ij}(\t) \v\xi(p)_{kl}(\t)>
&=& 2\v\t \d_{il} \d_{jk}                    \ , \nn
\ena
The index $p$ specifies the $p$-th matrix $M(p)_{ij}$ and
the white noise $\v \xi(p)_{ij}$.
$p$ runs from 1 to $N_0$, namely we consider $N_0$ matrices. The size of
all the matrices is assumed to be $N$.
The basic loop variable is typically of the form,
\EQ
tr(\Pi_\a M(p_\a)^{n_{p_\a}})
= tr (M(p_1)^{n_{p_1}}M(p_2)^{n_{p_2}}...M(p_\a)^{n_{p_\a}}...)  \ ,
\EN
The Langevin equation for these variables
takes the form similar to the one matrix model case. Namely, it includes
only linear terms and
bilinear terms of loop variables ( annihilation of a loop and simultaneous
creation of a pair of loops ). The correlation of the noise variables
is given by a linear combination of loops. It realizes the process where
various pairs of loops are annihilated simultaneously and a loop is created.
In this case, the colored noise correlation is still equivalent
to the insertion of matrices. Therefore,
if we identify the length of the loops by the power of
the matrices included in the loop variable, we may conclude that there exists
Virasoro constraint with respect to the loop length indices
in the loop variables ( string fields ).

In conclusion, we have shown that the Langevin equation ( or equivalently
the corresponding F-P hamiltonian ) for real symmetric
matrix models written by the loop variables
defines the time evolution in the ( non-critical ) non-orientable
string field theory at both
discretized and continuum levels. The partition
function for non-orientable strings satisfies the
Virasoro constraint at the equilibrium limit in both
discretized and continuum level. In the stochastic
quantization view point, since the fictitious time scaling dimension
is given by ${D\over 2} - 2 = {1\over 2} > 0$ for $c = 0$, we expect that
the discretized version of the loop space Langevin equation
for real symmetric matrix models
may provide a possible method for numerical calculation
of non-orientable 2D random surfaces to sum up the topologies of surfaces.

\vs{10}
\noindent
{\it Note}

After the completion of this work, we found the works, refs.~\cite{GG}
\cite{Dou}\cite{Mig}, where the second quantization of master fields are
discussed by stochastic quantization method. The expectation
value of master fields corresponds to eq. (9) in the present scheme.

\vs{10}
\noindent
{\it Acknowledgements}

The author would like to thank J. Ambjorn, H. B. Nielsen and
Y. Makeenko for valuable discussions and comments.
He also wishes to thank Y. Watabiki for discussions on the transfer matrix
formalism and all members
in high energy group at Niels Bohr Institute for hospitality.



\begin{thebibliography}{99}
%
%
\bibitem{SFT}
M. Kaku and K. Kikkawa, \PR{D10}(1974) 1110;1823;\\
W. Siegel, \PL{B151}(1985) 391;396;\\
E. Witten \NP{B268}(1986) 253;\\
H. Hata, K. Itoh, T. Kugo, H. Kunitomo and K. Ogawa,
\PL{B172}(1986) 186;195;\\
A. Neveu and P. West, \PL{B168}(1986) 192.
%
\bibitem{IK} N. Ishibashi and H. Kawai, \PL{B314}(1993) 190.
%
\bibitem{JR} A. Jevicki and J. Rodrigues, \NP{B421}(1994) 278.
%
%
\bibitem{Wata} Y. Watabiki, preprint, INS-1017,
hep-th/9401096; INS-1038, hep-th/9407058.
%
\bibitem{IIKMNS} N. Ishibashi and H. Kawai, \PL{B322}(1994)67; \\
M. Ikehara, N. Ishibashi, H. Kawai, T. Mogami, R. Nakayama
and N. Sasakura, KEK-TH-402, hep-th/9406207.
%
\bibitem{KKMW} H. Kawai, N. Kawamoto, T. Mogami and Y. Watabiki,
\PL{B306}(1993) 19.
%
\bibitem{DT} F. David, \NP{B257}(1985)45; \\
V. A. Kazakov, \PL{B150}(1985)282; \\
D. V. Boulatov, V. A. Kazakov, I. K. Kostov and A. A. Migdal,
\NP{B257}(1986)641; \\
J. Ambjorn, B. Durhuus and J. Fr${\dot o}$hlich, \NP{B257}(1985)433.
%
\bibitem{PW} G. Parisi and Y. Wu, Sci. Sin. {\bf 24}(1981)483.
%
%
\bibitem{JS} A. Jeviki and B. Sakita, \PR{D22}(1980) 467.
%
\bibitem{RS} E. Brezin and H. Neuberger, \PRL{65}(1990)2098; \\
\NP{B350}(1991)513. \\
G. R. Harris and E. Martinec, \PL{B245}(1990)384. \\
%
\bibitem{I} K. Ito, Proc. Imp. Acad. {\bf 20}(1944)519; \\
K. Ito and S. Watanabe, in Stochastic Differential Equations, ed. K. Ito
(Wiley, 1978).
%
\bibitem{GH} J. Greensite and M. B. Halpern, \NP{B211}(1983) 343.
%
\bibitem{Na} N. Nakazawa, \PTP{86}(1991)1053.
%
\bibitem{AJM} J. Ambjorn, J. Jurkiewicz and Y. Makeenko, \PL{B251}(1990)517.
%
\bibitem{FKN} M. Fukuma, H. Kawai and R. Nakayama,
\IJMP{A6}(1991) 1385; \\
R. Dijkgraaf, E. Verlinde and H. Verlinde, \NP{B348}(1991) 435.
%
\bibitem{DS} E. Br\'{e}zin and V. Kazakov, \PL{B236}(1990) 144;\\
M. Douglas and S. Shenker, \NP{B335}(1990) 635;\\
D. Gross and A. Migdal, \PRL{64}(1990) 127; \NP{B340}(1990) 333.
%
\bibitem{GG} R. Gopakumar and  D. J. Gross, hep-th/9411021.
%
\bibitem{Dou} M. R. Douglas, preprint, RU-94-81, hep-th/9411025.
%
\bibitem{Mig} A. A. Migdal preprint, PUPT-1509, hep-th/9411100.
%
%
\end{thebibliography}
\end{document}